# Displacement-agnostic coherent imaging through scatter with an interpretable deep neural network

**Authors:**

Yunzhe Li (emmal@bu.edu)

Shiyi Cheng (csy11@bu.edu)

Yujia Xue (yujiaxue@bu.edu)

Lei Tian[*] (leitian@bu.edu, Tel.: (617) 353-1334)

[*]**Corresponding Author**

**Affiliation:**

Department of Electrical and Computer Engineering, Boston University, Boston, MA 02215, USA

**Abstract**

Coherent imaging through scatter is a challenging task in computational imaging. Both model-based and data-driven approaches have been explored to solve the inverse scattering problem. In our previous work, we have shown that a deep learning approach can make high-quality and highly generalizable predictions through unseen diffusers. Here, we propose a new deep neural network (DNN) model that is agnostic to a broader class of perturbations including scatterer change, displacements, and system defocus up to 10X depth of field. In addition, we develop a new analysis framework for interpreting the mechanism of our DNN model and visualizing its generalizability based on an *unsupervised* dimension reduction technique. We show that our DNN can unmix the scattering-specific information and extract the object-specific information so as to achieve generalization under different scattering conditions. Our work paves the way to a highly robust and interpretable deep learning approach to imaging through scattering media.

**Introduction**

Imaging through scatter remains one of the most challenging tasks in computational imaging. The difficulty stems from the scattering process scrambling the object's spatial frequency information and generating a complex system matrix. As a result, computational retrieval of the object requires solving an ill-posed inverse problem based on the speckle measurements and a careful characterization of the random media[1,2]. Despite these challenges, many effective computational imaging techniques have been demonstrated for various applications, such as wavefront shaping[3], deep tissue imaging[4], and dynamic biological imaging[5].

In general, the coherent scattering process can be characterized by a linear transmission matrix (TM)[1,6]. This coherent TM establishes a one-to-one relation between the input and the output wavefronts. However, since the scattering is in general linearly shift-variant (LSV)[1,3,6], the complete characterization of the TM is often time-consuming due to the large size of the matrix[1]. Computational imaging techniques based on inverting the TM are susceptible to calibration errors, which may come from medium change and other perturbations to the system[7–10]. One useful simplification utilizing the memory effect[11,12] approximates the system to be linearly shift-invariant (LSI)[13,14]. Under this approximation, an invariant speckle intensity pattern only translates when shifting/rotating the incident beam over a small distance/angle[11,12]. This implies that under the *incoherent* imaging condition, the output speckle intensity is the convolution between the object's intensity distribution and the medium's speckle intensity point spread function (PSF)[13,15]. Based on this principle, 2D imaging through scatter can be achieved with a single-shot by either utilizing the pre-calibrated PSF[16] or solving a phase retrieval problem based on the autocorrelation of the output intensity[17].

Recently, deep learning (DL) has proven to be a powerful technique for solving highly ill-posed computational imaging problems[18]. In particular, deep neural network (DNN) models have been proposed to replace the standard TM that relates the output speckle patterns and the input objects, and shown superior performance over traditional methods[19–22]. Most importantly, the DNN models have shown to be resilient against various perturbations and instabilities of the scattering media. For example, DNN models for coherent imaging through multimode fibers can make robust predictions under temperature and mechanical instabilities[21,23,24]. In our previous work, we have shown that a DNN model for coherent imaging through scatter trained on a few thin diffusers can make high-quality predictions through unseen diffusers[25], indicating the model's robustness against medium perturbations. Specifically, during this study, we changed the diffuser placed in-between the object and the imaging optics while keeping the diffuser's location and the imaging system unchanged. In general, many other factors can perturb the scattering medium and thus affect the imaging performance.

In this work, we further consider the effect of axial displacements of the scatterer itself and the imaging optics. We demonstrate a more robust DNN model for coherent imaging through scatter that is agnostic to a broader class of perturbations. Generally speaking, axial displacements of both the scatterer and the imaging optics reduce the correlation of the speckle intensity measurements[26]. Recently, the 3D memory effect has been used to expand the imaging range and achieve extended depth-of-field (DOF) in incoherent imaging through scatter[27–30]. Coherent imaging under defocus is further complicated by diffraction effects. In Wu *et al.*'s work[31], 10X DOF improvement is achieved by incorporating defocus measurements in the training process for in-line holographic imaging in free space. In general, to train a robust DNN model against a broad class of perturbations, a diverse training dataset is needed to provide sufficient statistical information of the underlying process. As a result, we design our DNN training by incorporating the measurements taken from changing the scatterer, axially displacing the scatterer, and defocusing the imaging optics. Specifically, our training data includes speckle intensity patterns captured from four different diffusers at each training position; the training positions includes one diffuser position at 5X DOF displacement and two camera positions at $\pm$5X DOF defocus (Fig. 1(a)). Most importantly, we show that the trained DNN can make high-quality predictions *beyond* the training range which is across 10X DOF through previously unseen diffuser.

To achieve robustness to displacement and better generalizability, we propose a DNN model using a hybrid network structure to better model the shift-*variant* property of the imaging problem. Our network is built on the modified U-net structure in our previous work[25]. To improve the network's expressivity for modeling the SV properties, we add two fully-connected layers in the bottleneck of the network, as denoted by the transformation module in Fig. 1(b). This module takes input from the encoded features and transforms the 2D information to a 1D latent code, which is then fed into the decoder to reconstruct the 2D object. The operations on the latent code in the transformation module enlarges the effective receptive field of the DNN model, which in turn accounts for the shift *variance* of the system.

To *interpret* the working principle of our DNN model and better understand its *generalization* capability, we further develop a new analysis framework based on an *unsupervised* dimension reduction technique, UMAP[32]. In particular, our analysis provides several new insights into both the information contained in the raw speckle data, and the training and prediction processes. First, we show that directly decomposing the raw speckle intensity data onto a nonlinear manifold using the unsupervised technique already reveals scatter and displacement-specific information *without the need for any prior information nor supervised learning*. Second, by further analyzing the latent code in our DNN model, we show that the model reveals object-specific information and disentangles scatter and displacement information using the encoder path. The generalization of our DNN model is analyzed in two steps. First, we set up the *learned* latent manifold using the *learned* latent codes extracted by feeding only the *training* data to the *trained* network. Next, the *predicted* latent codes from the *unseen* speckle patterns under different scattering conditions are extracted and *projected* onto the learned manifold. We show that the predicted latent codes match well with the learned latent codes, which indicates that indeed the DNN model can generalize well to the unseen scattering cases. Finally, we further "dissect" the network and elucidate on the distinct functionalities of the encoder-decoder path and skip connections for performing the underlying imaging task. Our analysis shows that the encoder-decoder path is responsible to the generalization to *unseen* scatterers, whereas the skip connections mainly contribute to improve the reconstruction quality for seen scatterers.

In summary, we demonstrated a deep learning method for coherent imaging through scatter that is agnostic to scatterer change, displacements, and system defocus. We demonstrated robust predictions across 10X DOF by a new data acquisition procedure and a novel DNN model. We further developed a new analysis framework for revealing the information contained in the speckle data, interpreting the mechanism of the network, and visualizing the generalizability of the DNN model.

## Results

We develop a robust DNN model for coherent imaging through scatter that is agnostic to both change of scatterers and axial displacements of scatterer and the imaging optics, as summarized in Fig. 1.

*Experimental setup*

The imaging setup is shown in Fig. 2(a). A spatial light modulator (SLM) (Holoeye NIR-011, pixel size 8 um) was coherently illuminated by a collimated beam from a HeNe laser (632 nm, Thorlabs HNL210L). We used the SLM as a programmable amplitude-only object by placing two orthogonally oriented polarizers before and after. The SLM was relayed onto the camera (Thorlabs Quantalux, pixel size 5.04 µm) by a 4F system. Two lenses with focal lengths 200 mm (L1) and 125 mm (L2) were used to provide a 0.625 magnification. A 14 mm iris was placed at the pupil plane of the 4F system to control the speckle size. A diffuser was placed in between the SLM and L1. We placed five diffusers (Thorlabs N-BK7 Ground Glass Diffuser, 220 Grit DG10-220) on a filter wheel (Thorlabs FW1A) in order to capture data through different diffusers. Both the camera and filter wheel were attached to linear motion stages that can be moved axially. The initial positions for the camera $Z_{C10}$ and the diffuser $Z_{D0}$ were set at the back focal plane of L2 and 100 mm in front of L1, respectively. By controlling the motion stages, we captured speckle patterns from multiple combinations of diffuser/camera displacements and through five different diffusers. The intervals between the neighboring displacement positions for the diffuser and the camera are 1X DOF, which are set by the corresponding speckle sizes.

*System characterization*

The system was characterized by measuring the 3D speckle size[26]. We first captured a speckle intensity stack by moving the camera from 2 mm before $Z_{C10}$ to 2 mm after $Z_{C10}$ with step size 0.02 mm while fixing the diffuser at its initial position $Z_{D0}$. We then measured the 3D speckle size by calculating the 3D autocorrelation of the speckle intensity stack, as shown in Fig. 2(b). The experimentally measured lateral and axial speckle sizes at the sensor plane are 10.08 µm and 410 µm, respectively, which match well with the theory (see Materials and Methods). The axial speckle size at scatter plane is enlarged to 1.04 mm due to the system magnification. We used the respective axial speckle size to define the DOF in the object and image spaces.

*Data acquisition*

When taking the displacement measurements, we set the interval between two neighboring positions of the diffuser and the camera to be 1 mm and 0.5 mm, respectively. The total displacement range of the diffuser and the camera covers 10X DOF and 20X DOF, respectively. First, the diffuser was moved from the initial position $Z_{D0}$ towards L1 while taking data at 10 different positions ($Z_{D1}$ to $Z_{D10}$) with the camera being fixed at $Z_{C10}$, as shown in Fig. 2(c). Next, the camera was moved across 20 positions from $Z_{C0}$ to $Z_{C20}$ with the diffuser being fixed at $Z_{D0}$. Among the 20 camera positions, 10 positions ($Z_{C0}$ to $Z_{C9}$) were moved towards L2, the other 10 were moved away from L2.

We studied the statistical distribution of the measured speckle intensity data. As shown in Fig. 2(c), the estimated probability density functions (PDF) of the speckle patterns captured from different objects through different diffusers and/or at different displacement positions can all be fitted to the *same* speckle intensity distribution function[26] (see Materials and Methods). This highlights that all the *scattering-specific* and *object-specific* information are encoded in the higher order statistics and are hard to extract using standard statistical fitting techniques.

Our training data consists of 4200 image pairs, each of which consisted of the input object and the measured speckle pattern. The input objects contained 400 MNIST handwritten digits[33], 350 of which were used as the training objects. The speckle patterns for training were taken through four different diffusers, one diffuser position, and two camera positions. Specifically, the training diffuser position was 5X DOF from its initial position at $Z_{D5}$; the two training camera positions were ±5X DOF away from its initial positions at $Z_{C5}$, $Z_{C15}$, as shown in Fig. 3(a). The testing data consisted of 50000 speckle patterns taken under two different imaging conditions. In the first testing condition, we tested our network using speckle patterns through four seen diffusers (i.e. used for taking the training data) at 9 unseen diffuser positions and 18 unseen camera positions, and using both seen and unseen objects. In the second testing condition, we applied our network to speckle patterns from one *unseen* diffuser (i.e. never used during network training) at all diffuser and camera positions and using both seen and unseen objects.

*Network implementation*
We built a DNN shown in Fig. 1(b) to learn a statistical model relating the speckle patterns and the unscattered objects. The overall structure of the proposed DNN follows the U-net architecture with the modifications of replacing the convolutional layers with the dense blocks[25] and the additional fully connected layers at the "bottleneck" to perform latent code transformation. The input to the CNN is a preprocessed 128×128 speckle intensity. The input then goes through the "encoder path", which yields a stack of 4×4 latent code. The latent code includes case-specific information that encodes the displacement and diffuser parameter. Next, the latent code is flattened to a 1D vector, which is input to two fully connected layers and then reshaped to 2D. Together, this composes the latent code transformation module. This module enables transforming the case-specific information to meaningful object-specific features.

These operations on the latent code also enlarges the effective receptive field of the DNN model, which facilitates modeling the shift-variance effect of the imaging process. The decoder reverses the process that recombines the information into feature maps with gradually increased lateral details. Skip connections are used to transfer additional information from the encoder to the decoder without going through the bottleneck. The final output is a binary object prediction. Additional details of our DNN model and the benefit of the transformation module are discussed in Materials and Methods and Supplementary Materials Fig. S1 and Fig. S2.

*Experimental Results*

We demonstrated the robustness of our network against diffuser change as well as diffuser and the camera displacements on two types of experiments. All the experimental results were obtained using the *same single* network trained with the four diffusers at three different training positions.

**Results on axial displacements through seen diffusers.** We first tested our network using the speckle patterns from the same four trained diffusers at different unseen positions, as shown in Fig. 3. The testing objects consisted of both seen digits used for the training and unseen digits. The testing displacement positions for both the diffuser and the camera were up to 10X DOF, as shown in Fig. 3(a). The speckle patterns appear notably different when the diffuser or the camera is displaced over 1X DOF in the axial direction. Our DNN demonstrated the ability to make high-quality predictions at previously unseen positions across 10X DOF. Representative examples of the speckle and prediction pairs for both seen and unseen objects are shown in Fig. 3(b). We first show the results on diffuser displacements with the camera placed at its initial position $Z_{C10}$. On the left panel of Fig. 3(b), the speckle patterns and the prediction results are shown for the diffuser position at 1X, 3X, 7X and 10X DOF, respectively. Next, we present the testing results for camera displacements with the seen diffusers placed at $Z_{D0}$ on the right panel of Fig. 3(b). We show the speckle patterns and the network predictions when the camera was displaced by -9X, -7X, 2X, 10X DOF away from $Z_{C10}$. Additional prediction results through seen diffusers are provided in Supplementary Materials Fig. S3.

**Results on imaging through unseen diffusers across different displacements.** In the second experiment, we further tested our network using the speckle patterns obtained with the unseen diffuser and across a range of displacement positions, as shown in Fig. 4(a). We tested our network on both seen and unseen objects from the MNIST digit dataset. As summarized in Fig. 4(b), the left panel shows that the prediction results with the diffuser displaced at 1X, 3X, 7X DOF, respectively. The right panel shows the prediction results with the camera displaced at -9X, 0X, 7X DOF, respectively. By using an unseen diffuser, the problem becomes more challenging as the network needs to overcome both scatter and position perturbations. As a result, the performance degrades as compared to the seen diffuser case. Still, the main structure of the objects are accurately recovered across a range of tested displacement positions. Additional prediction results through unseen diffuser are provided in Supplementary Materials Fig. S4.

**Quantitative evaluation results on imaging through seen and unseen diffusers with different displacements.** Next, we quantify the prediction performance using the Pearson correlation coefficient (PCC), as summarized in Fig. 5. Our experimental results show consistent predictions on both seen and unseen objects for both seen and unseen diffusers. Accordingly, the two curves plotted in Fig. 5 were calculated by accumulating the statistics from all the objects (seen and unseen) on all four seen diffusers and the single unseen diffuser, respectively. The mean PCC at each position over all the predictions are marked by the cross markers for the seen diffuser and the circle markers for the unseen diffuser. Each error bar quantifies the standard deviation of the prediction results at each position. We observed that for the seen diffusers, the network performs the best at the trained positions. When the displacement increases, the PCC gradually decreases. The variations of the predictions, quantified by the standard deviation (std), also increases with the displacement distance. For the unseen diffuser, the overall performance drops slightly and is in quantitative agreement to that reported in our previous work[25]. Notably, the prediction results are less dependent on the positions. The mean and std of the PCCs remain consistent over the entire displacement range. Overall, our DNN showed the ability to make high-quality predictions against perturbations from camera and diffuser displacements. The degradation of the predictions as a function of displacement is *gradual*, as seen in Fig. 3(b). This shows the robustness of our DNN model under these physical perturbations. Additional quantitative evaluation on the prediction results are provided in Supplementary Materials Fig. S6.

*Analysis*

Next, we investigated the correlation across different speckle patterns imaged under different imaging conditions and further developed a framework to interpret the mechanism of how our DNN model generalizes over different scattering conditions. To do so, we used the state-of-the-art *unsupervised* dimension reduction technique, UMAP[32]. UMAP models the entire dataset into a single nonlinear manifold by learning the underlying topological structure contained in the high-dimensional data. In its simplest form, UMAP considers each data (e.g. an image) as a single vector in the learned manifold and models the entire dataset as a 2D (nonlinear) representation. We apply this technique to analyze both the raw speckle patterns and the DNN model's latent codes, and propose a procedure to visualize the training and prediction process.

**Raw speckle patterns contain scattering-specific information.** We analyzed the input data and the corresponding measured speckle patterns to discover the *intrinsic* correlations, as shown in Fig. 6(a). First, the UMAP learned manifold of the input object dataset is visualized as a 2D map. For better visualization, We randomly selected 4000 images from the same MNIST dataset as our input data. As clearly shown in Fig. 6(a)(i), this dataset naturally (i.e. without any supervision / labels for UMAP) clusters into 10 groups corresponding to the underlying 10-digit classes, each of which is marked by a different color. Here, each point (i.e. a vector) on this 2D map represents an input object image. This visualization shows that the raw input object *intrinsically* contains object-specific information in its image structure as expected. Next, we visualized the UMAP learned manifold from 9600 speckle patterns taken under 24 different scattering conditions, including 4 diffusers, each with 3 diffuser positions and 3 camera positions. Importantly, we observe that the learned manifold for the speckle patterns are clustered into 24 distinct groups according to the underlying scattering condition, while the object-specific information has been scrambled by the scattering process, as shown in Fig. 6(a)(ii). We label each scattering condition by (D#, P#), where D1 - D4 indexes the four different diffusers and P1 - P6 indexes six different position, corresponding to $Z_{c5}, Z_{c15}, Z_{D5}, Z_{c1}, Z_{D3}, Z_{D9}$. We observe that speckles measured from the same diffuser at different positions do *not* form an apparent "super cluster". These results show that the speckle patterns captured under the same scattering condition contain intrinsic correlations; the speckle patterns become more decorrelated as the scattering condition changes. These observations match well with our previous study based on the classical Pearson correlation analysis[25]. Here, by using a more advanced dimension reduction technique, we show that the raw speckle patterns contain scattering-specific information that can be revealed *without the need for any supervised learning procedure*.

**Interpreting the mechanism of the DNN model's generalization.** Next, we develop a novel procedure to interpret the working principle of our DNN model and its generalization capability to different scattering conditions. Our main idea is to analyze the training and prediction processes and quantify the underlying information content using UMAP. To do so, we take the following two-step process. First, we set up the *learned latent space* by the *trained* DNN model, which will be used as the global "coordinate system" to quantify the information content in both the training and the prediction. Specifically, we fed each training data into the trained network and extracted the corresponding latent code. We then used all the latent code from the entire training dataset and set up the learned latent space using UMAP. Second, we *projected* the latent code extracted from the *testing* data under different conditions to the learned latent space (the coordinate system) and visualized the discrepancy between the learned and the predicted latent codes in order to assess the DNN model's generalization capability. In the following, we first discuss our analysis results on unseen objects under the same training scattering conditions and show that our DNN model can reveal object-specific information and disentangles scattering-specific information using the DNN model's encoder path. Next, we discuss the testing results underlying different scattering conditions and demonstrate our DNN's generalization capability to different scattering cases. Finally, we dissect our DNN model and discuss the distinct functionalities provided by the encoder-decoder path and the skip connections for solving this inverse scattering problem.

**The DNN model reveals object-specific information.** We first visualized the speckle patterns used for the training that include 12 different scattering conditions (i.e. four diffusers at one diffuser position and two camera positions) by UMAP in the 2D map in Fig. 6(b)(i), which is termed the *training input manifold*. As expected based on our previous analysis, 12 distinct clusters are formed matching the underlying scattering condition. Next, we fed all the training speckle patterns to our trained DNN and extracted the *learned* latent codes, which are then used to compute the *learned latent manifold* by UMAP. In Fig. 6(b)(v), the learned latent space is visualized as a 2D map. Importantly, it contains 10 clusters based on the corresponding digit label instead of the scattering conditions. This shows that the trained network learns to distill object-specific features and "unmix" the scattering effects.

Next, we projected the testing speckle patterns captured from unseen objects and under the same 12 scattering conditions onto the existing training input space under the same (nonlinear) UMAP transformation. As shown in Fig. 6(b)(ii), the projection aligns well with the existing training input manifold and the corresponding clusters, which further indicates that speckles captured with a given scattering condition are correlated regardless of the input objects. Next, we fed all the testing speckle patterns to the trained DNN and extracted the *predicted* latent codes. Finally, we projected the *predicted* latent codes onto the previously *learned* latent manifold, as shown in Fig. 6(b)(vi). The predicted latent code clusters align well with that from the training data. This result indicates that our DNN can reveal object-specific information under the same scattering conditions from unseen speckle patterns.

**Interpreting the DNN model's generalizability to different scattering conditions.** We discuss the analysis results from different scattering conditions in two cases. In the first case, we analyzed the testing data from different displacements through the same four training diffusers, including four unseen diffuser positions and four unseen camera positions. After projecting the input speckle patterns onto the previously established training input manifold, the clusters no longer align with the existing input manifold, as shown in Fig. 6(b)(iii). In the second case, we analyzed speckle patterns captured from the *unseen* diffuser with 4 diffuser positions and 4 camera positions. As shown in Fig. 6(b)(iv), the projected clusters significantly differ from the training input manifold. In both cases, it shows again that speckles captured from different scattering conditions are decorrelated. Although the manifold learned by UMAP is *nonlinear*, Fig. 6(b)(iii) and Fig. 6(b)(iv) can be intuitively interpreted as follow: given the "coordinate system" set up by the training speckles, the testing speckles can no longer be represented by any single cluster ("axis"). This is because speckle patterns from different cases exhibit different features so that their corresponding 2D representations are far apart. Specifically, combining with our previous analysis, the underlying scattering condition dictates the unique features in the speckle patterns, and will differentiate them from other cases. This elucidates on the challenge for deep learning to generalize over different scattering conditions.

The next step is to project the predicted latent codes extracted from the testing data onto the learned latent manifold under the same UMAP transformation. As shown in Fig. 6(b)(vii) and Fig. 6(b)(viii), for both cases under different scattering conditions, the predicted latent-code clusters align well with that of the training data. The good alignment between the predicted and learned latent manifolds illustrates our DNN model's ability to generalize in terms of diffuser displacements, camera displacement, and change of diffusers. Artifacts including mixing across different clusters are also observed as compared to the original learned latent manifold. These artifacts become more obvious for the *unseen* diffuser case (Fig. 6(b)(viii)). Fundamentally, this is because the learned encoder is trained based on the training data distribution, which may not sufficiently capture the testing data distribution.

**The encoder-decoder path and the skip connections provide distinct functionalities.** While the concept of latent code is originally established on the auto-encoder network (i.e. an encoder-decoder network without skip connections)[34], we adapt the same concept to analyze our modified U-net. To justify our approach, we compared three different network structures using the weights directly loaded from our trained network. Notably, we found that the encoder-decoder path and the skip connections provide distinct functionalities for inverting the coherent speckles.

In Fig. 7(a)(i), the modified U-net used in this work and a few representative prediction results on seen and unseen diffusers as shown as the benchmark. Next, in Fig. 7(a)(ii) we blocked the skip connections of the trained network to effectively construct a plain encoder-decoder network (see details in Materials and Methods). The corresponding prediction results showed that only minor blurring is resulted in the reconstruction. We further quantitatively evaluate the predictions from the plain encoder-decoder network in Fig. 7(b) using the same testing data as Fig. 5. Notably, the performance on the *unseen* diffuser across all positions remain consistently around 0.6 and is similar to that from Fig. 5 using the full U-net structure. Slight performance drop was observed on the *seen* diffuser cases with reduced mean PCC and increased std. Additional quantitative evaluation using Jaccard index is provided in Supplementary Materials Fig. S7.

Finally, in Fig. 7(a)(iii) we blocked the information flow from the latent code to the first layer of the decoder. The corresponding prediction results showed severe degradation. Overall, this study shows that the information used for reconstruction is primarily extracted from the encoder-decoder path. In particular, the encoder-decoder path plays the key role in generalizing to new scattering conditions. Adding the skip connections helps restoring high-resolution features and slightly improves the prediction results for the seen diffuser case. Because of this, we can confidently extend the latent code concept to analyze the information learned by our network without missing much from those flow through the skip connections.

Empirically, we also found that adding the skip connections helps preventing overfitting. To show this, we optimized the hyper-parameters of the plain encoder-decoder network [in Fig. 7(a)(ii)] and trained it from scratch using random initial weights on the same training dataset. The prediction results showed that the average PCCs for the seen diffuser cases improved slightly to around 0.9, while the PCCs on the unseen diffuser dropped to around 0.5, as summarized in Supplementary Materials Fig. S8. Next, we trained a plain encoder-decoder network using the same structure as our network and further initialized using the encoder-decoder path weights from our trained network. The validation loss gradually increased using this training scheme as shown in Supplementary Materials Fig. S9, which indicated that continued training based on previously learned weights also cannot improve generalization performance of the plain encoder-decoder network. Overall, this study shows that the plain encoder-decoder network tends to overfit to the *seen* diffuser dataset and does *not* generalize well to *unseen* diffusers. This issue was effectively overcome by the skip connections in our network.

**Discussion**

In this paper, we presented a new deep learning framework for coherent imaging through scatter and pushing the robustness against scatterer displacement and imaging optics defocus. We developed a new analysis framework for revealing the information contained in the speckle dataset, interpreting the mechanism of our DNN, and visualizing the generalizability of the DNN model. We established the importance of the encoder-decoder path for the DNN model's generalization to untrained scatterers, as well as how the skip connections help predicting overfitting to seen scattering conditions.

We demonstrated that our DNN model is agnostic to scatterer changes, scatterer displacement and camera defocus for coherent imaging through scatter. By improving the data acquisition strategy and improving the network structure, the generalization capability may be further improved, which will be explored in our future work. These promising results show that deep learning can robustly solve challenging inverse problems of coherent imaging through complex media under various perturbations.

Our analysis framework shows that the speckle patterns intrinsically carry scatter/displacement information. After training, the encoder can unmix the scattering and distill the object-specific information. Our analysis framework allows us to visualize the DNN's generalizability under different imaging conditions. This provides a powerful tool to explore the underlying correlations within the data and to interpret the learning mechanism of the DNN model. The caveat of using a nonlinear dimension reduction process, like UMAP, is that the traditional distance measures for linear spaces, such as the Euclidean norm, can no longer be used[32]. This poses challenges to quantify the discrepancy between the learned and the predicted distributions as well as to provide a unique inverse transform from the latent space to the input[32], both of which will be investigated in our future work.

Here, to simplify the problem we have focused on establishing an understanding of the mechanism of network generalization against scatterer and displacement changes while using a simple dataset containing only 10 classes of handwritten digits. This approach has the benefits of facilitating direct visualization of the learned latent space into a small number of visually distinguishable clusters using the dimension reduction technique. However, the simplicity of the object dataset inevitably limits the network's generalizability for predicting more complex structures, which requires increased diversity of the training objects, as shown in our previous work[25]. In general, it has been recently shown that the network's generalizability against object variations can be improved by increasing the information capability, i.e. entropy, of the training data[35]. In our work, the information content in the input data is visualized by the dimension reduction technique. Importantly, our results show that for systems involving complex transformations, such as scattering, directly measuring the information content in the raw input data may not provide sufficient information for generalization. Instead, we develop a latent code analysis framework for understanding this challenging computational imaging problem. We envision a comprehensive analysis framework developed by analyzing the information content of both the network input and the latent space may provide additional insights into improving and interpreting generalization against *both system perturbations and object variations*, which will be considered in our future work.

Our current network performs a deterministic inversion and makes a "point estimation" of the underlying object. For solving complex inverse problems, it has been shown that quantifying the uncertainty using a probabilistic perspective of deep learning can further enhance the reliability and interpretability of the prediction. For example, a variational auto-encoder framework has developed to visualize the prediction variations from inverting the speckle data[36]. A full Bayesian neural network framework has been developed to quantify the uncertainties due to both model and data variations when solving a phase retrieval problem[37].

Incorporating such uncertainty quantification technique may be particularly useful to fully understand the "deep" correlations contained in the coherent speckle data, which will be explored in our future work.

**Materials and Method**

**Theoretical 3D speckle size.** The speckle size in 3D was calculated by computing the 3D autocorrelation of a speckle intensity stack[26]. The theoretical lateral speckle size in free space propagation geometry is defined by the full width at half maximum (FWHM) of the normalized autocorrelation function along the lateral direction and is $1.0\lambda\frac{z}{D}$, where $\lambda$ is the wavelength of the coherent source, z is the distance between the scattering surface and observation region, D is the diameter of the scattering spot. The axial speckle size is defined by the FWHM of the normalized correlation function along the axial direction and is $7.1\lambda(\frac{z}{D})^2$, which also defines the system's DOF[26]. According to this, the theoretical DOF of our system at the object side is 0.92 mm and at imaging side is 0.38mm. The theoretical lateral speckle size at the camera side is 5.56 µm. Our experimentally calculated speckle intensity autocorrelation is shown in Fig. 2(b), which shows that the lateral speckle size is 10.08 µm. Due to under-sampling by the camera, the experimentally measured lateral speckle size is larger than the theoretical value.

**Speckle statistical distribution.** We investigated the statistical distribution of speckle intensity patterns of our dataset. For fully developed speckles, the intensity follows the negative exponential distribution. For $N$ incoherently summed independent speckles, the PDF is a Gamma density function[26].

$$P(I) = \frac{N^N I^{N-1}}{\Gamma(N)\bar{I}^N}\exp\left(-N\frac{I}{\bar{I}}\right)$$

In practice, the probability density function (PDF) was estimated from the normalized intensity histogram of experimentally measured speckle patterns. In Fig. 2(c), we show several representative cases that cover all the cases we studied, including the speckle patterns from different objects, different scatters, different camera positions, and different diffuser positions. Applying the theoretical PDF in MATLAB to match the experimental measurement, $N$ was estimated to be 1.8, which is consistent with our under-sampled imaging condition. All speckle intensity patterns approximately follow the same distribution.

**Data preprocessing.** The SLM input and the camera measurements are collected in pairs for generating the dataset. The central 512×512 SLM pixels were used as the object; the corresponding central 512×512 camera pixels were used as the speckle intensity input for our DNN. The objects displayed on the SLM were 8-bit grayscale images from the MNIST handwritten digit. Due to the computation and memory limitations, all input and output images were down-sampled from 512×512 pixels to 128×128 pixels by taking the average within each 4×4 neighboring pixels. Details on the comparisons about different downsizing methods are given in Supplementary Materials Fig. S2. For both training and testing, intensity outliers were removed by histogram clipping. The speckle images were then normalized between 0 and 1 by dividing each image by its maximum. Although we display grayscale images on the SLM, our DNN was designed to make binary predictions.

**Dimensionality Reduction and Data Visualization.** UMAP was used to model the data on a high-dimensional manifold with a graph structure to reduce the dimension for visualization. Once a large dataset without labels is fed into UMAP, the algorithm adaptively outputs an unsupervised transformation mapping between the high-dimensional dataset and a low-dimensional representation[32]. Consider the entire speckle dataset as a $N \times 128 \times 128$ matrix, where $N$ denotes the number of images. For perform dimensionality reduction analysis, we first preprocessed the data by reshaping the matrix into a $N \times 128^2$ matrix. Next, we compute the training input manifold as a 2D representation using UMAP. To analyze the testing data, we use the transform method in UMAP to directly project preprocessed 2D testing input onto the training input manifold. For latent code analysis, the learned manifold was first computed from the latent code by feeding the entire training dataset into the trained network and extracting the code from the bottleneck layer. Next, we used the transform method in UMAP to project the testing data's latent code to the previously established learned manifold.

**Neural network implementation.** The input to the CNN is a preprocessed 128×128 speckle intensity. Next, the input goes through the "encoder path", which consists of 4 dense blocks connected by a max pooling layer for down-sampling. Each dense block consists of 2 layers, in which each layer performs batch-normalization (BN), the rectified linear unit (ReLU) nonlinear activation, and convolution (conv) with 3 filters. The intermediate output from the encoder has small lateral dimensions (4×4), but encodes rich information along the "depth". A transformation module is then concatenated to the encoder, consisting of a flatten layer that outputs a 1-D latent vector, 2 fully connected layers with ReLU activation and a constant reshaping layer that transforms back to a 2D feature map. Next, the low-resolution feature maps go through the "decoder path", which consists of 4 additional dense blocks connected by up-sampling followed by convolutional layers. The information across different spatial scales are tunneled through the encoder-decoder paths by skip connections to preserve high-frequency information. After the decoder path, an additional convolutional layer with sigmoid followed by the last layer produces the network output. The last layer is designed to solve a pixel-wise binary prediction problem. The CNN makes decisions on if an object is present. The CNN training was performed on BU Shared Computing Cluster with one GPU (NVIDIA Tesla P100) using Keras/Tensorflow. Each CNN was trained with 100 epochs by the ADAM optimizer for up to 3 hours. The learning rate of $10^{-4}$ was used. Once the CNN was trained, each prediction was made in 0.0156 s. More details about the neural network implementation can be accessed in Supplementary Materials Fig. S1.

**Blocking the skip connections or latent code path.** To block the information flow of the skip connections or the latent code, we modified the layer weights using the following procedures. Since each skip connection is implemented by a merge layer (concat) that concatenates the feature maps from the encoder layer to the matching decoder layer followed by a 2D convolutional layer (2D conv), we extracted the weights of 2D conv layer and set those for bridging the information from the encoder layer to be zero. By doing so way, each concat layer effectively only passes the decoder feature maps to the next layer while blocking the feature maps from the skip connections. In a similar fashion, the latent code information can be blocked by setting the weights of 2D conv layer immediately after the bottleneck layer to be zero. By doing so, the first decoder layer always outputs zero feature maps regardless of the latent code. To validate this approach, we also constructed a plain encoder-decoder network without the skip connections, while keeping the rest of hyper-parameters the same as our trained network. We then passed the corresponding weights from the trained network to the plain encoder-decoder network. Next, we performed predictions using both the plain encoder-decoder network and our network with blocked skip connections on the same testing data, and showed identical prediction results, as shown in Supplementary Materials Fig. S5.

**Acknowledgements**
We thank George Barbastathis and Mo Deng for insightful discussions, Boston University Shared Computing Cluster for proving the computational resources.

**Competing interests**
The authors declare that there are no conflicts of interest related to this article.

**Contributions**
Y.L. and L.T. conceived the idea. Y.L. built the setup, conducted the experiments, devised the deep learning model, and established the data analysis method. S.C. refined with the deep learning model and latent code interpretation. Y.X. contributed to the data censoring, results evaluation. L.T. supervised the whole project. All authors discussed the results and contributed to the writing of the manuscript.

**Funding**
National Science Foundation (1813848).

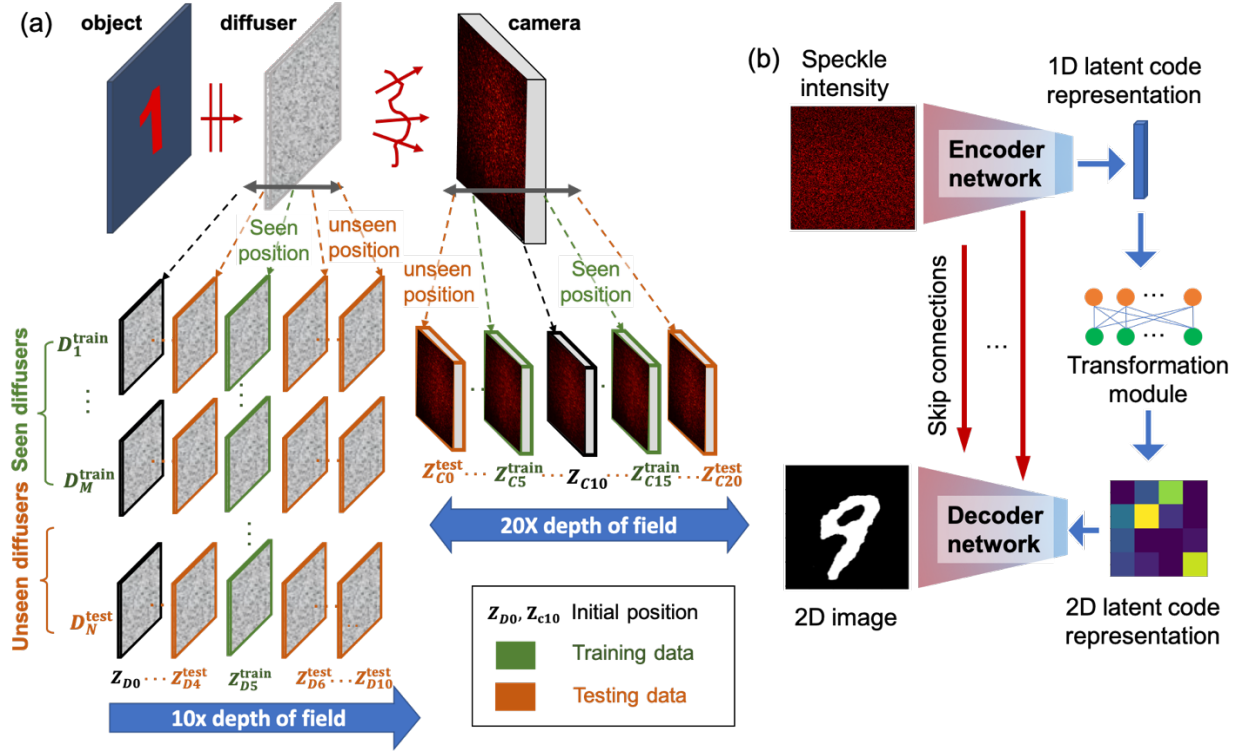

**Fig. 1. Overview of our deep learning-based approach to achieve generalization for coherent imaging through scatter.** **(a)** Our coherent imaging model and data acquisition approach to obtain a diverse dataset. The dataset including scatterer changes, sensor and scatterer displacement over 10X DOF. Speckle patterns from training diffusers at training positions are used to train the DNN. Others are used as testing data. **(b)** We implemented a DNN structure including a transformation module to achieve generalizability.

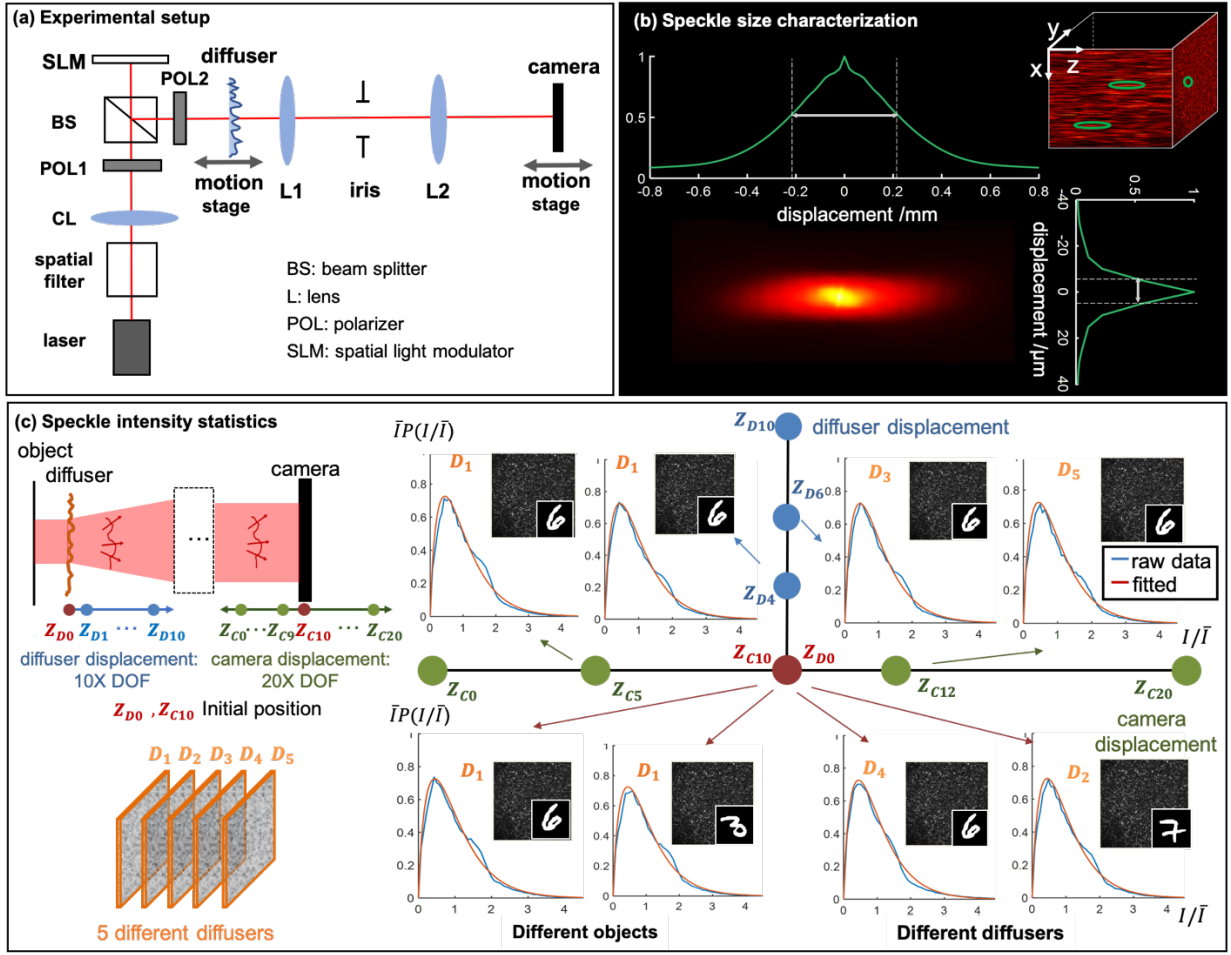

**Fig. 2. Experimental setup and speckle characterization. (a)** The experiment setup for coherent imaging through scatter. The SLM is used as the amplitude-only object. Both the diffuser and the camera are placed on motion stages to control the axial displacement. **(b)** The 3D speckle size is characterized by calculating the speckle intensity stack's autocorrelation. **(c)** The raw speckle intensity distributions $P(I/\bar{I})$ approximately follow the same probability density distribution regardless of the input object, diffuser, camera position $Z_{Ci}$, and diffuser position $Z_{Di}$.

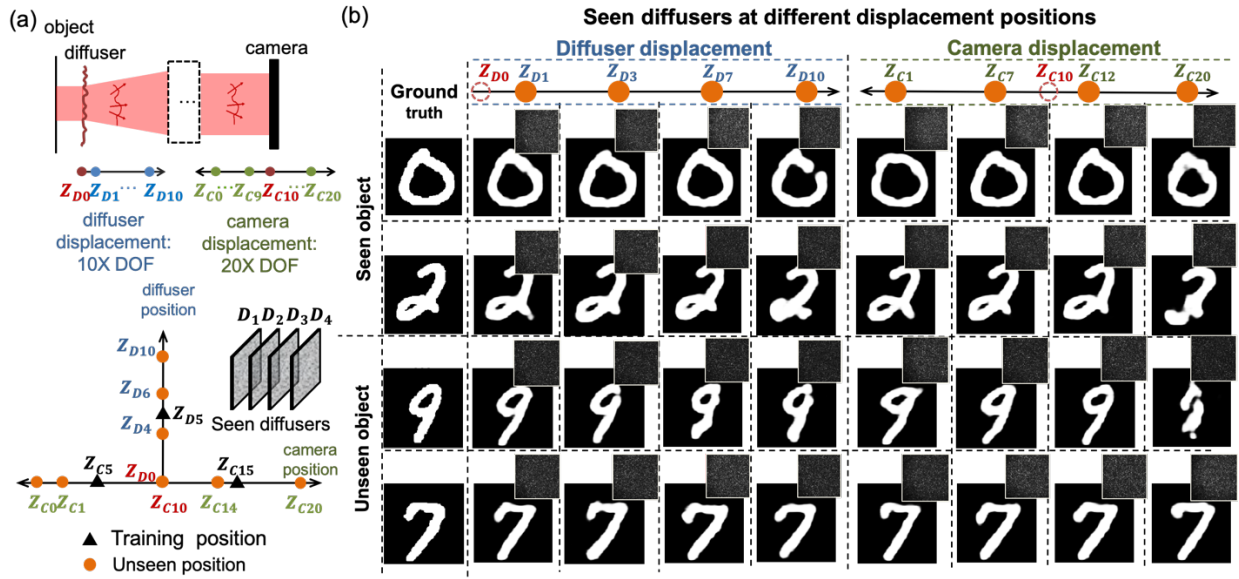

**Fig. 3. Data acquisition and results on coherent imaging through seen diffusers at unseen displacement positions. (a)** A summary of the training and testing dataset. The training data are captured through four different training diffusers at three training positions. All the rest are testing positions. **(b)** Representative testing results at different diffuser displacement positions (Left panel) and camera displacement positions (Right panel) for both seen objects (Row 1 and Row 2) and unseen objects (Row 3 and Row 4).

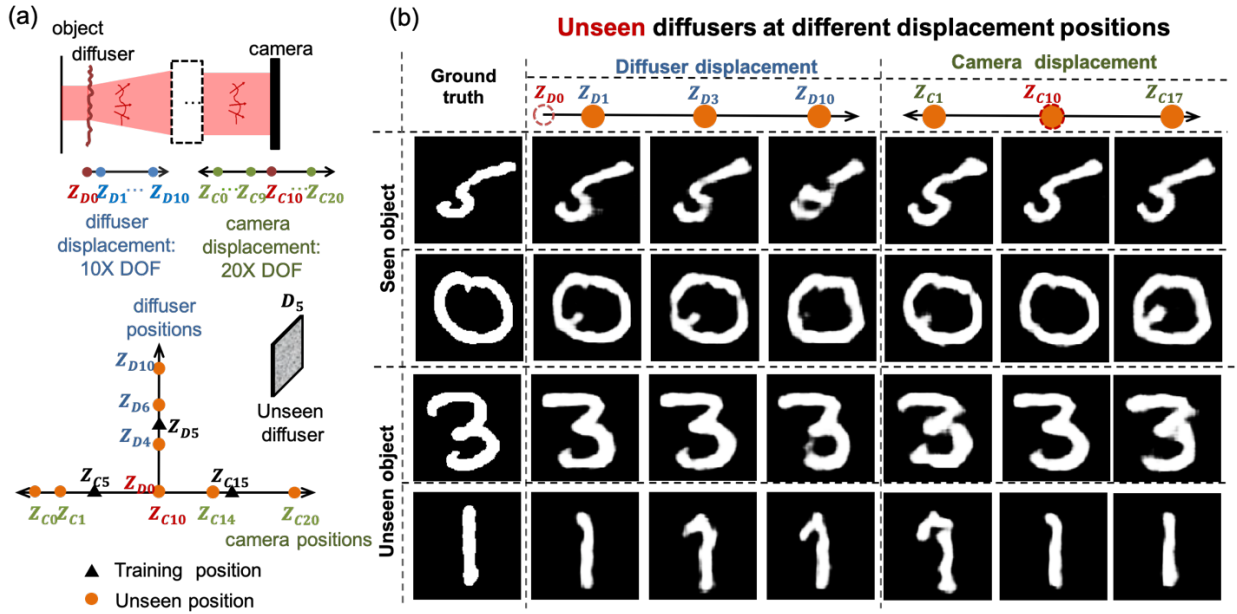

**Fig. 4. Data acquisition and results on coherent imaging through seen diffusers at unseen displacement positions. (a)** A summary of the training and testing dataset. Training data are the same as the data in the first experiment. Testing data is from the unseen diffuser $D_5$ at different positions. **(b)** Testing results through unseen diffuser $D_5$ at different diffuser displacement positions (Left panel) and camera displacement positions (Right panel) for both seen objects (Row 1 and Row 2) and unseen objects (Row 3 and Row 4).

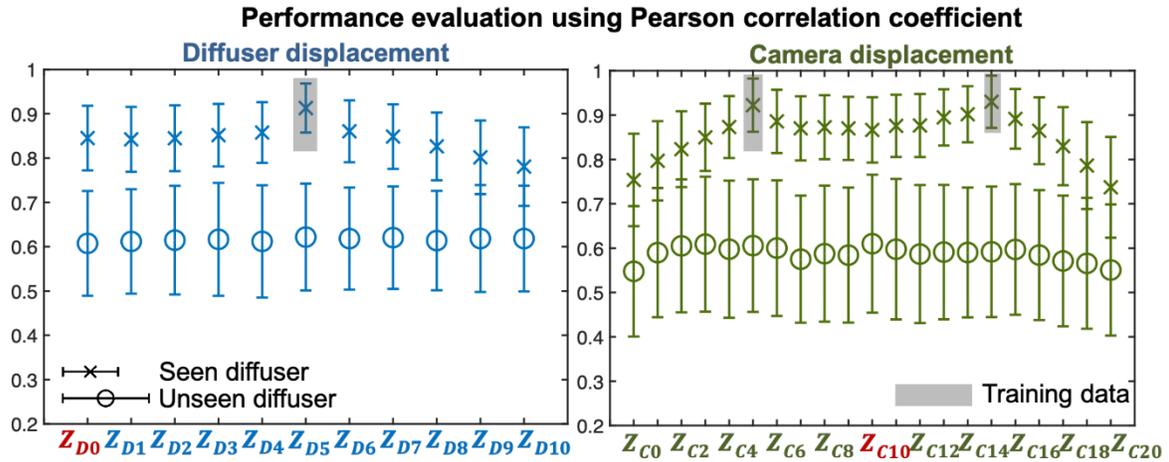

**Fig. 5. Quantitative performance evaluation of DNN prediction results.** Each cross (seen diffuser) or circle (unseen diffuser) marker represents the mean Pearson correlation coefficient (PCC) of the predictions at each position. Each error bar quantifies the standard deviation of the prediction results at each position. The training positions are marked by the grey boxes.

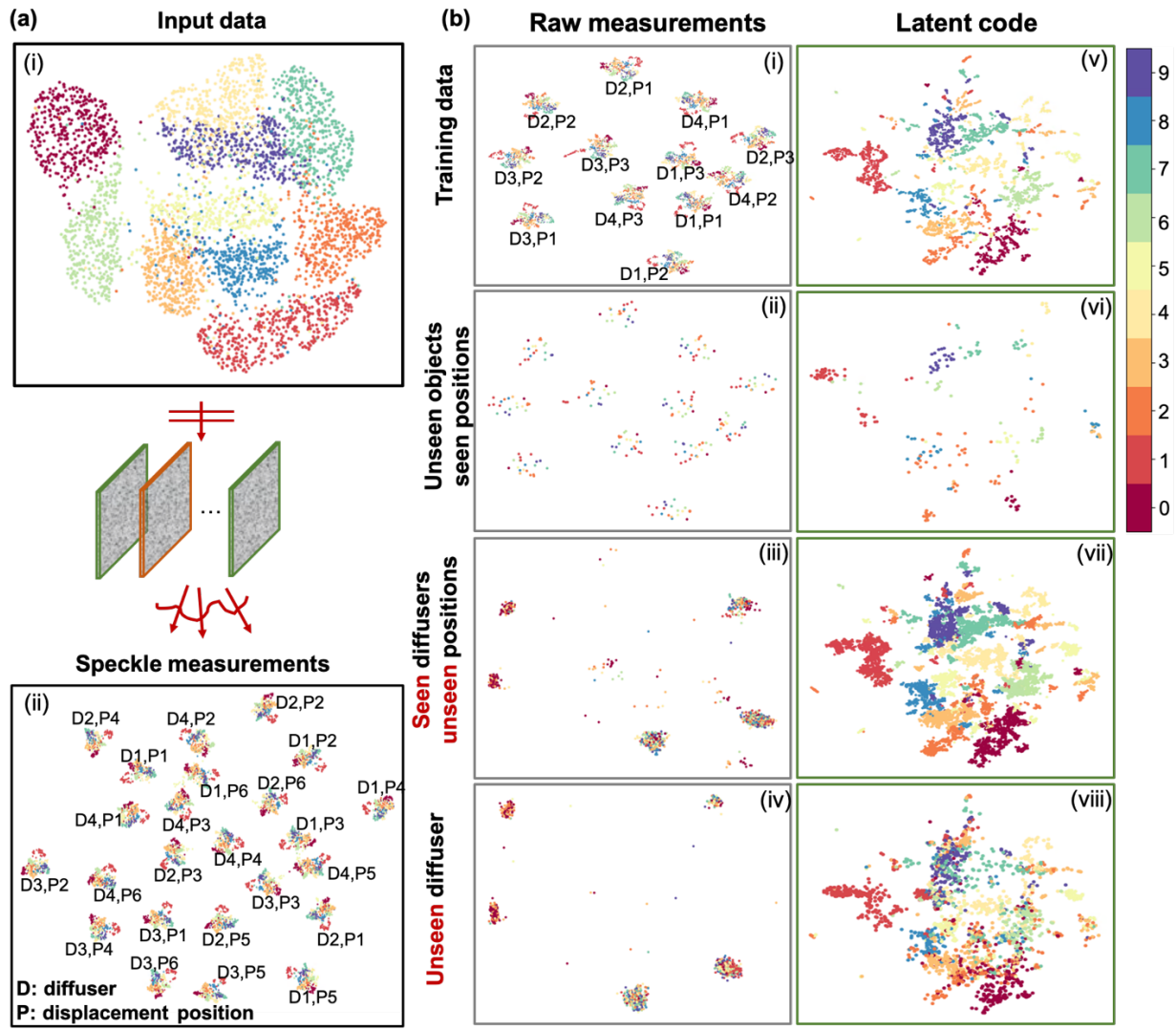

**Fig. 6. Data and latent space analysis based on dimension reduction. (a)** UMAP based visualization of the input data and speckle measurements. (i) The manifold of the input data shows 10 clusters matching the underlying 10 digits. (ii) The manifold of the speckle patterns forms clusters based on the underlying scattering conditions, in which the indices of the diffuser and the position are marked by D# and P#, respectively. **(b)** Network analysis during training and making predictions. (i) The training input manifold computed from the speckles captured under 12 scattering conditions correspondingly forms 12 distinct clusters. (ii-iv) The testing data under different imaging conditions are projected onto the same training input manifold. (v) The learned latent manifold is computed from the leaned latent code of the training data. (vi-viii) The corresponding *predicted* latent code for testing data under different imaging conditions are projected onto the learned latent manifold.

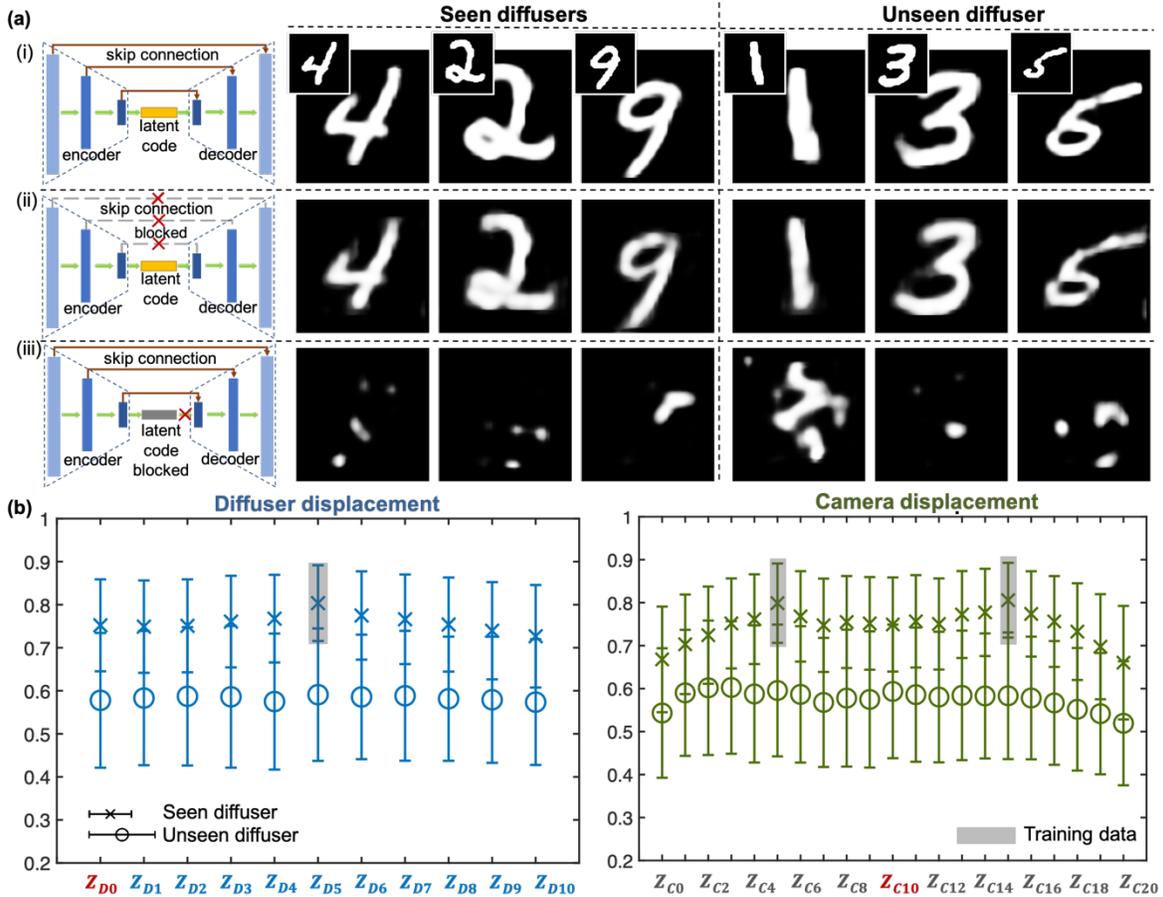

**Fig. 7. Analysis of the roles of the encoder-decoder path and the skip connections. (a) An overview** of three different network structures under study and their representative prediction results on seen and unseen diffusers. (i) Our DNN model contains both an encoder-decoder path and skip connections. (ii) The encoder-decoder network with the skip connections being blocked. (iii) The skip-connection only network with the latent code layer being blocked. **(b)** Quantitative evaluation of the performance from the network in Fig. 7(a)(ii) without skip connections. Each cross (seen diffusers) or circle (unseen diffuser) marker represents the mean PCC of the predictions at each position. Each error bar quantifies the standard deviation of the prediction results at each position. The training displacement positions are marked by the grey box.